# Global optimization of proteins using a dynamical lattice model: Ground states and energy landscapes

F. Dressel[1,2] and S. Kobe[1]

[1]Institut für Theoretische Physik, [2]Biotechnologisches Zentrum, Technische Universität Dresden, D-01062 Dresden, Germany

## Abstract

A simple approach is proposed to investigate the protein structure. Using a low complexity model, a simple pairwise interaction and the concept of global optimization, we are able to calculate ground states of proteins, which are in agreement with experimental data. All possible model structures of small proteins are available below a certain energy threshold. The exact low-energy landscapes for the trp cage protein (1L2Y) is presented showing the connectivity of all states and energy barriers.

The most challenging problem in quantitative biology today would be the response to the question, just how the spatial structure of a protein is encoded in its sequence of amino acids. The huge number of degrees of freedom per amino acid with respect to its spatial structure is reflected in the high computational complexity in continuous space protein models. Their variants with space discretization are called high coordination lattice models [1]. Alternatively, simple lattice models are widely investigated, based on the assumption that the positions of amino acids are represented by vertices of a regular lattice (see e.g. [2]). In these cases, there are only z-1 possibilities for the position of the next amino acid in a chain, where the coordination number z is 4 for the square lattice and varies between 6 and 12 for the standard 3d lattices. Using simple lattice models, one may study the qualititative behaviour of protein folding and dynamics. However, these models restrict the finding of the real three-dimensional structure because of rigidity in the underlying crystalline (predefined) lattice. We propose a dynamical lattice model (DLM), which belongs to another class of low-coordination number models: the so-called reduced ($\varphi$–$\psi$) or ($\alpha$-$\tau$) models. Its introduction is motivated by the request to combine low computational complexity with the possibility of an adequate description of the spatial structure. Other investigations on the basis of such reduced discrete state models are done by Park and Levitt [3] without considerati-on of amino-acid specificity, whereas Zhang et al. [4], [5] have used an amino-acid specific discrete state model.

The protein backbone consists of repeating sequences of nitrogen (N), carbon ($C_\alpha$) and carbon (C) atoms, which on their part, form the backbone of the amino acids. In our model, the atomic distances in the protein backbone are fixed in a common way: C-N = 1.32 Å, C-C = 1.53 Å, N-$C_\alpha$ = 1.47 Å. Additionally, the angles between every three of these atoms are fixed, too: N-$C_\alpha$-C = 110º, $C_\alpha$-C-N = 114º, C-N-$C_\alpha$ = 123º. Because the $C_\alpha$ atom of the succeeding amino acid lies in the plane $C_\alpha$-C-N, its position respective to the preceding amino acid is determined by three angles: $\varphi$, $\psi$, $\omega$. Applying biological constraints, one of them ($\omega$) can be fixed to 180º. In real proteins, the remaining angles $\varphi$ and $\psi$ are correlated and different from one amino acid to another. We extract $q$ relevant correlations per amino acid by a cluster analysis over angle pairs ($\varphi,\psi$) from a data set consisting of 403 proteins [6]. This leads to a mean value of $q_{av}$ = 3.65. A cluster analysis with a coarser resolution ($q_{av}$ = 3.0, see Table 1) provides reasonable information about structure with less computational effort.

**Table 1:** Angle pairs ($\varphi,\psi$) obtained from a cluster analysis with $q_{av}$ = 3.0 and hard-core diameter $d_{SC}$ (in Å) of the 20 amino acid (aa) side chains.

| aa | ($\varphi,\psi$) | $d_{SC}$ |
|---|---|---|
| A | (-135,150), (-65,-35), (-75,140) | 4.0 |
| C | (60,35), (-110,140), (-75,-30), (-120,45) | 4.4 |
| D | (55,35), (-95,140), (-110,60), (-70,-30) | 4.4 |
| E | (-105,135), (-70,-35) | 4.7 |
| F | (-115,140), (-70,-35), (-115,40) | 5.1 |
| G | (95,-165), (85,10), (-125,170), (-70,-30) | 3.6 |
| H | (60,35), (-115,135), (-80,-25) | 4.9 |
| I | (-125,130), (-65,-45), (-95,-5), (-95,130) | 4.9 |
| K | (-105,140), (-70,-30) | 5.1 |
| L | (-100,135), (-70,-35) | 4.9 |
| M | (-110,135), (-70,-35) | 4.9 |
| N | (55,40), (-105,130), (-80,-20) | 4.5 |
| P | (-60,-25), (-75,165), (-65,140) | 4.4 |
| Q | (60,40), (-70,-30), (-110,140) | 4.8 |
| R | (-70,-35), (-110,135) | 5.2 |
| S | (-135,150), (-70,-25), (-80,150) | 4.1 |
| T | (-130,160), (-100,135), (-95,-5), (-65,-40) | 4.5 |
| V | (-130,145), (-100,130), (-115,0), (-65,-40) | 4.7 |
| W | (-110,140), (-60,-45), (-80,-25) | 5.4 |
| Y | (-115,135), (-75,-30) | 5.2 |

The side chains are modeled by hard spheres with volumes corresponding to the Van-der-Waals volumes of the real side chains. The centres of the spheres lie on the straight line $C_\alpha$-$C_\beta$ (in the case of glycine a virtual $C_\beta$ is assumed). Their distances from the $C_\alpha$ correspond to their radii ($d_{sc}/2$, see Tab. 1). The $C_\beta$ atoms are placed at one corner of a tetrahedron built up by $C_\alpha$ at the centre (see Fig. 1), which is selected according to the CORN law [7]. To account for volume exclusion of the protein backbone, we choose an additional distance constraint for every two $C_\alpha$ atoms $i$ and $j$: $r_{ij} \geq 3$ Å.

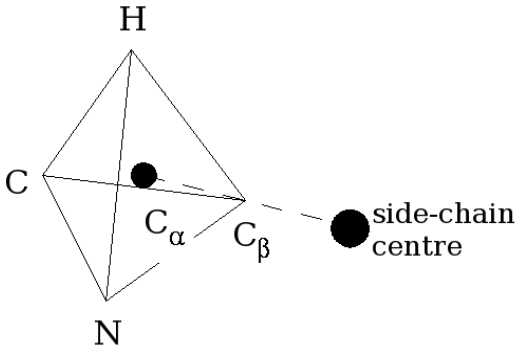

**Fig. 1:** The orientation of the N, $C_\beta$, H, C atoms and the center of the side chain is determined by a tedrahedron with $C_\alpha$ at its centre.

On the basis of the proposed model, we have taken into account 20 different types of amino acids with interaction constants $e_{\mu\nu}$ between amino acids μ and ν. The strength of the interaction energy $E_{ij}(r_{ij})$ is modified according to

$$E_{ij}(r_{ij}) = e_{\mu\nu}\{\tanh[(8.0-r_{ij})/2]/2+0.5\}, \quad (1)$$

where $r_{ij}$ (in Å) is the distance between the $C_\beta$'s of the $i^{th}$ and $j^{th}$ amino acid along the chain. This is a smooth approximation to a stepwise function with cutoff at 8.0 Å [8].

Our goal in the first stage was to find the ground state, which minimizes the total energy of a DLM protein with $n$ amino acids

$$E_{gs} = min \sum_{i,j}^{n} E_{ij} \quad (2)$$

with $i,j \geq 1$, $i > j-1$. This was done using an algorithm of discrete optimization. In the second stage, we extended the method to access the exact low-energy landscape.

The method is based on *branch-and-bound* and was applied first to a magnetic Ising system with frustration by Kobe and Hartwig [9]. The strategy of branch-and-bound is to exclude as many states with high energy values as possible in an early stage of calculation. Using this method, we are able to calculate the exact ground state of the DLM, which has altogether about $3^{30}$ different states for $n \approx 30$.

In the language of Statistical Physics, the proposed model is equivalent to a "mixed $q$-state Potts glass" with 20 different Potts elements (amino acids), where $q$, in general, is different for each of them. The meaning of a 'Potts state' corresponds to an angle pair (φ−ψ).

First, we tested the model and the optimization algorithm for small proteins. We found, e.g., a right-handed α-helix for the DLM ground state of α-helix protein (Protein Data Bank (PDB) entry 1AL1 [10] with the chain length $n = 13$. As an example for a non-helical stucture, the DLM ground state of *compstatin* is shown in Fig. 2 in comparison with the corresponding structure of the PDB, both drawn using PyMol

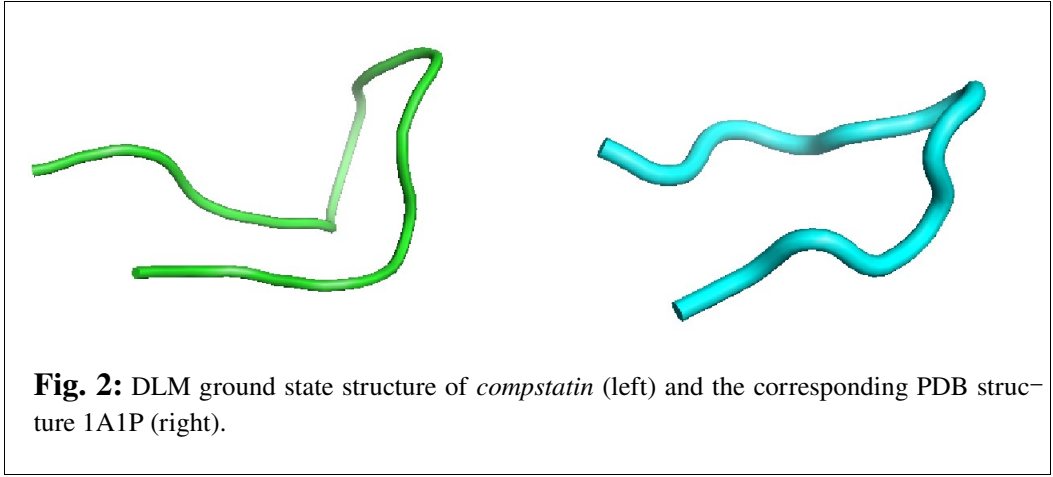

**Fig. 2:** DLM ground state structure of *compstatin* (left) and the corresponding PDB structure 1A1P (right).

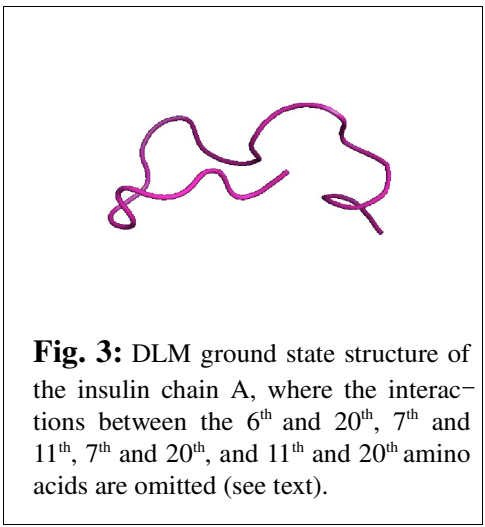

**Fig. 3:** DLM ground state structure of the insulin chain A, where the interactions between the 6[th] and 20[th], 7[th] and 11[th], 7[th] and 20[th], and 11[th] and 20[th] amino acids are omitted (see text).

[11].

The result for the chain A of insulin with the sequence GIVEQCCTSICS-LYQLENYCN is shown in Fig. 3. In nature, the A chain is coupled with the B chain by disulfide bonds, which are tied at the cysteines at positions 7 and 20 in the A chain. But in our example, the A chain is computed in absence of its companion. Consequently, the non-bonded cysteines in our model would form disulfide bonds within the chain, causing distortions with respect to the expected spatial structure. To avoid these distortions, we have switched-off the respective interactions between the 6[th] and 20[th], 7[th] and 11[th], 7[th] and 20[th], and 11[th] and 20[th] amino acids. Herewith, we have obtained in the DLM ground state a structure, which forms a helix between position 1 and 7 in agreement with the helical structure part in the PDB. There are remaining deviations from the real structure, which are probably due to the neglected interaction with the B chain.

The sequence length $n$ of computable peptides can be increased by restriction of the regarded interactions. When one considers only the interactions up to the $l^{th}$ neighbor of each amino acid along the chain, the summation in eq. (2) has to be restricted to $j-l \leq i < j-1$. With this procedure, we handle the Alzheimer disease peptide 1AML choosing $l = 7$. One of the few representative energetically low-lying states is shown in Fig. 4. This structure agrees well with the according PDB secondary structure, which assigns a helix in the middle part (amino acid 14 - 24) and one at the end (amino acid 32 - 35) as well as some bends between. These structural elements have been found in the calculated conformation at right sequential positions. In the considered case of a relatively stretched protein, the tertiary structure shows small deviations due to neglected long-range interactions along the chain, binding the peptide together. In general, the restriction of $l$ cannot be applied to more compact proteins.

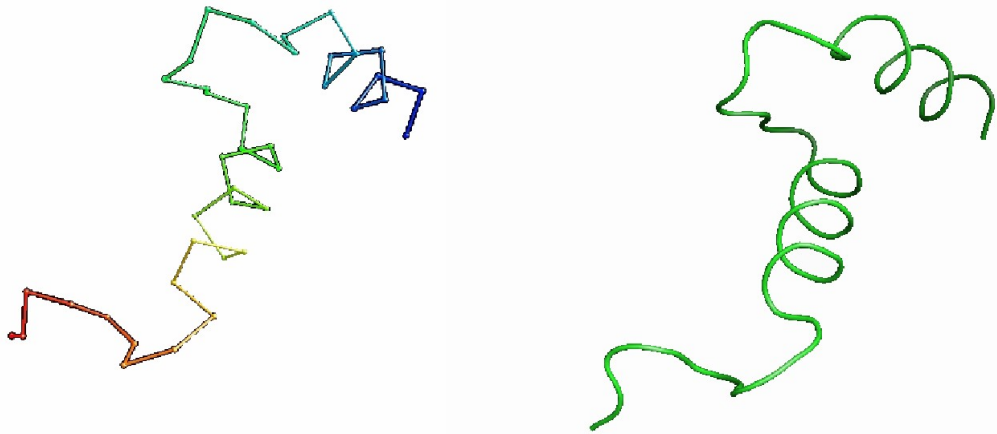

**Fig. 4:** One representative low-lying DLM structure of the Alzheimer's disease amyloid peptide (A4). Only interactions of amino acids up to the seventh neighbour along the chain are taken into account; $C_\alpha$ ball-and-stick representation (left), PyMol representation (right).

For a more quantitative comparison of the DLM ground-state structures with experimental results, we have calculated the RMSD (root mean square deviation) with respect to the $C_\alpha$ atoms (Tab. 2). Obviously, it cannot be the aim of a coarse-grained model with only few degrees of freedom per amino acids to compete with the accuracy of modern programs of structure prediction. On the other hand, the relevance of the DLM structures are underlined by relatively good RMSD values. In many cases, the secondary structure agrees with PDB

**Table 2:** RMSD (in Å, calculated according to [12]) for comparison of some DLM ground states with experimental results of the PDB [10]. $RMSD_{est}$ is an estimated greatest lower limit of RMSD (see text).

| PDB code | name | n | RMSD | $RMSD_{est}$ |
|---|---|---|---|---|
| 1AL1 | $\alpha$-helix protein | 13 | 1.98 | 0.94 |
| 1A1P | compstatin (Fig.2) | 14 | 2.46 | 2.20 |
| 1AKG | conotoxin | 17 | 3.21 | 2.83 |
| 1L2Y | trp cage (Fig. 5) | 20 | 5.90 | 2.58 |
| 1L2Y | trp cage ($q_{av}$ = 3.0) | 20 | 3.73 | |
| 1D9J | cecropin-magainin hybrid | 20 | 4.58 | 3.10 |
| 1B19:A | insulin A (Fig. 3) | 21 | 5.65 | |
| 1G04 | sheep prion segment | 26 | 8.60 | 4.93 |
| 1ANP | atrial natriuretic peptide variant ($q_{av}$ = 3.0) | 28 | 6.67 | 6.28 |
| 1AML | Alzheimer A4, $l = 7$ (Fig. 4) | 40 | 5.99 | 5.64 |

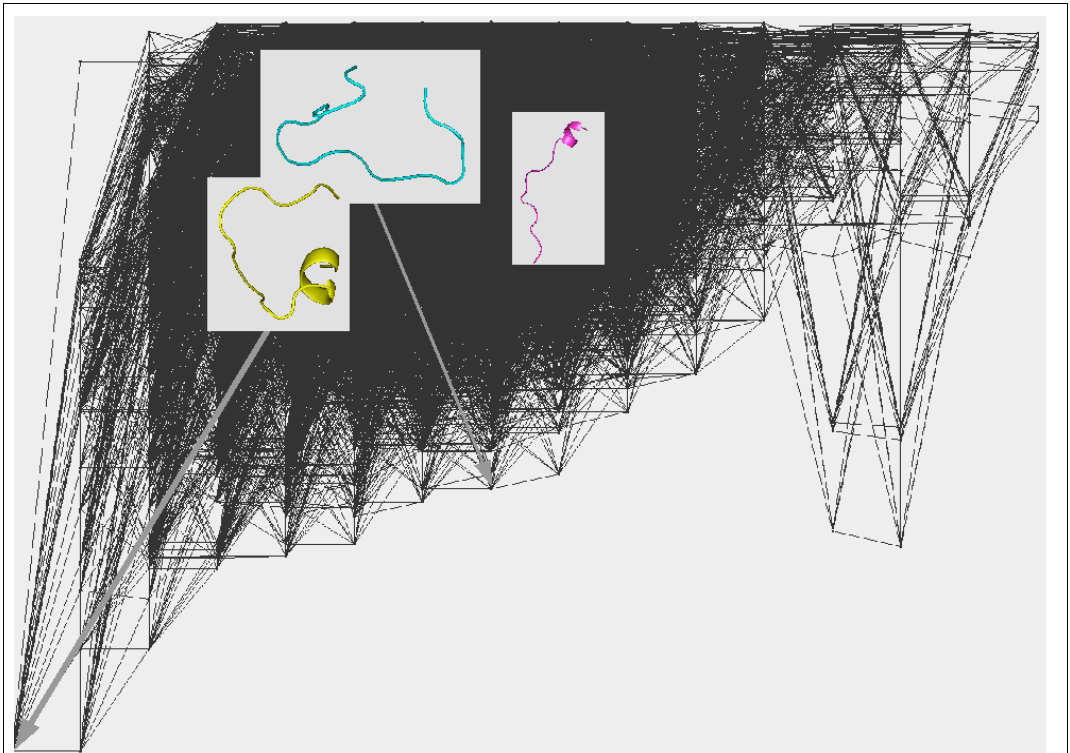

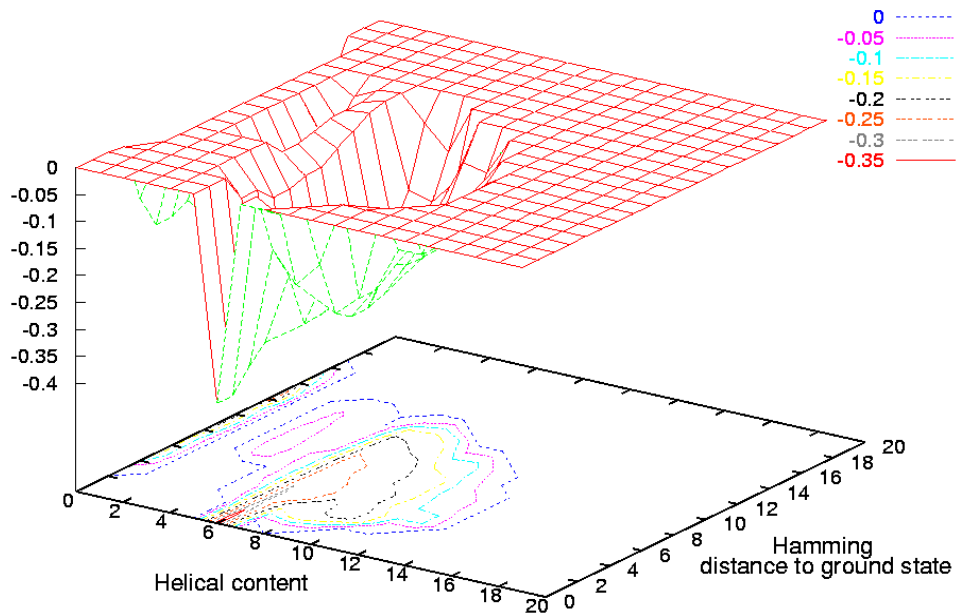

**Fig. 5:** Energy landscape of trp cage protein (1L2Y). Energy vs. Hamming distance with respect to the ground state. All possible transitions are dented by lines. Characteristic structures of some regions are shown in the insets (above). Energy barriers (in a.u.) vs. helical content (number of amino acids, which adopt helical configurations) and Hamming distance with respect to the ground state (below).

[10]. We have estimated RMSD$_{est}$ as the greatest lower limit of RMSD by a simple stochastical fit to the PDB.

The relative low complexity of the DLM is of particular importance for investigations of the energy landscape of proteins. So, it is of great interest to study post-optimal states, which can be candidates for misfolded conformations. For this purpose, the used *branch-and-bound* algorithm is extended in a second step in such a way, that *all* energetically low-lying conformations are exactly calculated [13]. An advantage of this procedure is, that it provides the complete information of the model protein: energy, conformations and all possible transitions between them.

The trp cage protein (1L2Y, sequence NLYIQWLKDGGPSSGRPPPS) is used as a common protein to test models, cf. [14]. In Fig. 5 (above) the energy landscape of 1L2Y is shown. The energies of all 8884 states below a bound energy of 0.68 $E_{gs}$ are plotted against the Hamming distance from the ground state. The Hamming distance denotes the number of amino acids, whose 'Potts state' differs from its 'Potts state' in the DLM ground state. Any two DLM states, which differ only in the 'Potts state' of one of the amino acids, are connected by lines. The conformations of selected states are given in the insets showing different helical contents in different regions of the landscape. The ground state is two-fold degenerated. Its structure is similar to the PDB entry, starting with a α-helix and ending with a lefthanded tail, which is one of the characteristics of this protein. In yet another inset, the structure of one post-optimal state is shown. A third inset shows a typical random-coil-like structure for that region of the energy landscape, in which it is situated. In Fig. 5 (below) the height of energy barriers is plotted in dependency of the helical content and the Hamming distance with respect to the ground state.

There are three attracting basins: the ground state basin and two competitors, where the ground state is dominating. The latter is in good agreement with experiments showing a fast folding time and with other recent simulations. The appearance of competing basins is not in contradiction to a funnel-like energy landscape [15] because their reachability is negligible. There are two reasons: firstly, the number of states and the possible transitions into these basins are low, and secondly, the energy barriers are high. .

Summarizing, we have used a minimalistic amino-acid specific reduced discrete state model for proteins. Contrary to models using simple regular lattices, the discrete lattice sites in the proposed model are dynamically created in dependence upon the amino acid sequence. As a result, one can obtain representative spatial structures of proteins. We have shown that most of the DLM ground state structures correspond to a high degree with the measured native structures. This is reflected in the agreement of secondary structure and in part, tertiary structure.

The DLM achieves good computability because of its relatively low complexity. It is shown that complete low-energy landscapes of realistic three-dimensional protein structures can be investigated. On this basis, the DLM opens the possibility to study the dynamics of the system.

We thank K. Binder, V. Lankau, W. Naumann, A. Schneider and J. Weißbarth for discussions and L. Huber for critical reading of the manuscript. S. K. is grateful to A. Hartwig for three decades of continuous cooperation and discussions on global optimization.